\documentstyle[12pt,epsf]{article}
\relax
\begin{document}
\begin{titlepage}
\vfill
\hskip 4in BNL-62549

\hskip 4in ISU-HET-95-9

\vskip 1.2in
\begin{center}
{\large \bf The Decays $K \rightarrow \pi \pi \nu \overline\nu$ within the 
Standard Model}\\

\vspace{1 in}
{\bf L.~S.~Littenberg$^{(a)}$ and G.~Valencia$^{(b)}$}\\
{\it  $^{(a)}$ Physics Department,
               Brookhaven National Laboratory,  Upton, NY 11973}\\
{\it  $^{(b)}$ Department of Physics,
               Iowa State University,
               Ames IA 50011}\\
\vspace{1 in}
\end{center}
\begin{abstract}

We study the reactions $K \rightarrow \pi \pi \nu \overline{\nu}$ 
within the minimal standard model. We use isospin symmetry to  relate the 
matrix elements to the form factors measured in $K_{\ell 4}$. 
We argue that these modes are short distance dominated 
and can be used for precise determinations of the CKM parameters 
$\rho$ and $\eta$. Depending on the value of the CKM angles we find 
branching ratios in the following ranges:  
$B(K_L \rightarrow \pi^+ \pi^- \nu \overline{\nu}) = [2-5] \times 10^{-13}$; 
$B(K_L \rightarrow \pi^0 \pi^0 \nu \overline{\nu}) = [1-3] \times 10^{-13}$; 
$B(K^+ \rightarrow \pi^+ \pi^0 \nu \overline{\nu}) = [1-2] \times 10^{-14}$. 
We also discuss a possible $CP$-odd observable.

\end{abstract}
\end{titlepage} 
\clearpage

Rare kaon decays have long been recognized for their potential 
to measure the CKM matrix parameters $\rho$ and $\eta$ as well 
as for their sensitivity to certain types of new interactions beyond 
the minimal standard model. Rare decays involving a lepton 
anti-lepton pair are predominantly mediated by four fermion 
operators that can be thought of as the product of a hadronic and 
leptonic currents. In this way it is possible to relate the 
hadronic matrix element to a measured 
semi-leptonic decay and avoid the uncertainties that are inherent 
to purely hadronic decays. This is particularly true for processes 
in which the leptons are neutral since they do not have long 
distance contributions from radiative kaon decays \cite{review}. 

The short distance analysis for $|\Delta S| =1$ transitions into a 
$\nu \overline\nu$ pair has been carried out in detail before. 
The dominant contribution arises from penguin and box diagrams 
with intermediate top and charm quarks. It can be written in the 
form of an effective Lagrangian \cite{short,buras}:
\begin{equation}
{\cal L} = {G_F \over \sqrt{2}}{\alpha \over 2 \pi \sin^2\theta_W}
\biggl[V_{cs}^*V_{cd} \overline{X}(x_c,y_\ell)+V_{ts}^*V_{td}
X(x_t)\biggr] 
\overline{s}\gamma_\mu(1-\gamma_5)d \overline\nu \gamma^\mu 
(1-\gamma_5) \nu + {\rm H. c.}
\label{shortd}
\end{equation}
where the dependence on the charm-quark, top-quark and 
tau-lepton masses in terms of 
$x_i=M_i^2/M_W^2$ and $y_\ell= m_\ell^2/M_W^2$ is contained in the 
functions:
\begin{equation}
X(x_t)={x_t\over 8}\biggl[{x_t+2\over x_t-1}+3{x_t-2\over
(x_t-1)^2}\log x_t\biggr], 
\label{xofxt}
\end{equation}
and $\overline{X}(x_c,y_\ell)$. The function $\overline{X}(x_c,y_\ell)$ is 
the analogue of Eq.~\ref{xofxt} for a charm-quark intermediate 
state. In this case, however, the tau-lepton mass dependence is important 
as are the QCD corrections. This function cannot be written as compactly as 
Eq.~\ref{xofxt} but it can be found in Ref.~\cite{buras}.

To compute the differential decay rate for the process $K \to 
\pi \pi \nu \overline\nu$ we need to compute the matrix element of 
the hadronic current $\overline{s}\gamma_\mu(1-\gamma_5)d$
between the kaon and two pions states. In this note we will 
extract the current matrix element from the one measured in $K_{\ell 4}$ 
using isospin symmetry. 

The standard analysis of $K_{\ell 4}$ proceeds in terms of the form 
factors defined by \cite{pais}:
\begin{eqnarray}
{\big< \pi^+(p^+)\pi^-(p^-) \big | \overline{s}\gamma_\mu \gamma_5 u \big |
K^+(k) \big>} &=& 
- -{i \over M_K}\biggl[F(p^+ +p^-)_\mu  + G (p^+ -p^-)_\mu 
\nonumber \\ && +
R (k-p^+-p^-)_\mu  \biggr] \label{klppff} \\
{\big< \pi^+(p^+)\pi^-(p^-) \big | \overline{s}\gamma_\mu  u \big | K^+(k) 
\big>} &=& 
{H \over M_K^3}\epsilon_{\mu\nu\alpha\beta}k^\nu (p^+ +p^-)^\alpha 
(p^+ -p^-)^\beta  
\end{eqnarray}
The contribution of the form factor $R$ to $K_{\ell 4}$ is suppressed by the 
lepton mass, and $R$ does not contribute to $K\to \pi \pi \nu \overline{\nu}$. 
The form factors determined in $K_{e4}$ decays \cite{klfexp} 
have been found to depend on the $\pi-\pi$ invariant mass only. 
Theoretically, one expects these form factors 
to depend on all the kinematical invariants of the reaction, and this 
is found in a $\chi PT$ calculation \cite{kltheory}. The dependence 
of the form factors on invariants other than $M_{\pi\pi}$ may lead 
to interesting interference effects in the reactions 
$K \to \pi \pi \nu \overline{\nu}$, but we defer this discussion to a future 
publication. With this caveat we proceed to use the form factors 
measured in $K_{e4}$ in terms of the variable 
$q^2=((p^++p^-)^2-4m_\pi^2)/4m_\pi^2$ and the $\pi-\pi$ scattering 
phase shifts $\delta^I_J$ \cite{klfexp}: 
\begin{eqnarray}
F&=&f_s(0)(1+\overline\lambda q^2)e^{i\delta^0_0} \nonumber \\
G&=&g(0)(1+\overline\lambda q^2)e^{i\delta^1_1} \nonumber \\
H&=&h(0)(1+\overline\lambda q^2)e^{i\delta^1_1} .
\label{ffexp}
\end{eqnarray}
The following constants have been measured 
(we use $\sin\theta_c=0.22$) \cite{klfexp}:
\begin{eqnarray}
f_s(0) &=& 5.59 \pm 0.14 \nonumber \\
g(0) &=& 4.77 \pm 0.27 \nonumber \\
\overline\lambda &=& 0.08 \pm 0.02  \label{ffvalues} \\
h(0) &=& -2.68 \pm 0.68 \nonumber
\end{eqnarray}
The current matrix element that we need may be extracted from these 
measurements in the following way: 
when the two pions are in an $I=0$ state, 
\begin{equation}
{\big< \pi^+(p^+)\pi^-(p^-) \big| \overline{s}\gamma_\mu (1-\gamma_5) d \big|
K^0(k) \big> } = 
{\big< \pi^+(p^+)\pi^-(p^-) \big| \overline{s}\gamma_\mu (1-\gamma_5) u \big|
K^+(k) \big> } ;
\end{equation}
and when they are in an $I=1$ state, 
\begin{equation}
{\big< \pi^+(p^+)\pi^-(p^-) \big| \overline{s}\gamma_\mu (1-\gamma_5) d \big|
K^0(k) \big> } = -
{\big< \pi^+(p^+)\pi^-(p^-) \big| \overline{s}\gamma_\mu (1-\gamma_5) u \big|
K^+(k) \big> } .
\end{equation}
Using this we find that
\begin{eqnarray}
A(K^0 \to \pi^+ \pi^- \nu_\ell \overline{\nu}_\ell) 
&=& -{G_F \over \sqrt{2}M_K}{\alpha \over 2\pi\sin^2\theta_W}A^2\lambda^5 
X(x_t) \hat{\lambda}_t \overline{\nu} \gamma^\mu (1-\gamma_5) \nu 
\label{koamp}\\
&& \biggl[F(p^++p^-)_\mu - G(p^+-p^-)_\mu + i {2 H \over M_K^2}
\epsilon_{\mu\nu\alpha\beta}k^\nu p^{+\alpha}p^{-\beta}\biggr]\nonumber 
\end{eqnarray}
where we have introduced the notation
\begin{equation}
\hat{\lambda}_t \approx 1-\rho - i \eta + {(1-\lambda^2/2)\over A^2 \lambda^4}
{\overline{X}(x_c,y_\ell)\over X(x_t)} 
\equiv \rho^0_\ell -\rho - i\eta
\end{equation}
and we use the Wolfenstein parameterization of the CKM matrix. 
{}From this we obtain\cite{cptinv}:
\begin{eqnarray}
&&A(K_L \to \pi^+ \pi^- \nu_\ell \overline{\nu}_\ell) 
= -{G_F \over M_K}{\alpha \over 
2\pi\sin^2\theta_W}A^2\lambda^5 X(x_t) \overline{\nu} \gamma^\mu (1-\gamma_5)
\nu \label{klamp} \\
&& \times \biggl[F(\rho^0_\ell -\rho)(p^++p^-)_\mu 
+i \eta  G(p^+-p^-)_\mu + i {2 H \over M_K^2}(\rho^0_\ell -\rho)
\epsilon_{\mu\nu\alpha\beta}k^\nu p^{+\alpha}p^{-\beta}\biggr].\nonumber
\end{eqnarray}
For our numerical estimates we use $\lambda=0.22$ and $V_{cb}=0.041$ 
(therefore $A\approx 0.85$). 
Integration over phase space yields the branching ratio
\begin{equation}
B(K_L \to \pi^+ \pi^- \nu \overline{\nu})= \sum_\ell\bigl[
6.1 (\rho^0_\ell -\rho)^2 + 0.9 \eta^2 \bigr]
\times 10^{-14}
\label{sdbr}
\end{equation}
The two terms in this expression come from the contributions of 
the $F^2$ and  $G^2$ terms in the squared matrix element. 
The first term corresponds to an $s$-wave, $I=0$ $\pi^+ \pi^-$ pair, 
whereas the second term corresponds to a $p$-wave, $I=1$ $\pi^+ \pi^-$ pair. 
The contribution of the $H^2$ term to $B(K_L \to \pi^+ \pi^- \nu 
\overline{\nu})$ integrated over phase space has the same $\rho$ 
dependence as the first term in 
Eq.~\ref{sdbr}, but is much smaller. Unlike $K_{\ell 4}$, where it is 
possible to reconstruct all the momenta, in $K \to \pi \pi \nu \overline{\nu}$ 
only the pion momenta can be reconstructed. This reduces the number of 
interference terms that can actually contribute to any observable in 
these reactions. With the momentum dependence of the form factors that 
we are using, only one interference term is potentially interesting. 
The $F-G$ interference gives rise to a $CP$-odd $E_{\pi^+}-E_{\pi^-}$ 
asymmetry in the kaon rest frame. We find for the integrated asymmetry 
\begin{eqnarray}
A_{CP}&\equiv&{1 \over \Gamma_{K_L}}\int d\Gamma(K_L \to \pi^+ \pi^- \nu 
\overline{\nu}) {\rm sign}[k\cdot(p^+ -p^-)] \nonumber \\
&\approx& 3.5 \times 10^{-15}\eta \sum_\ell(\rho^0_\ell -\rho)
\label{asym}
\end{eqnarray}
In a similar manner we find:
\begin{equation}
B(K_L \to \pi^0 \pi^0 \nu \overline{\nu}) = 3.2\times 10^{-14}\sum_\ell 
(\rho^0_\ell -\rho)^2 , 
\label{sdbrnn}
\end{equation}
reflecting the fact that the two neutral pions cannot be in an $I=1$ state; 
and also:
\begin{equation}
B(K^+ \to \pi^+ \pi^0 \nu \overline{\nu}) = 2.3\times 10^{-15}\sum_\ell 
\bigl[(\rho^0_\ell -\rho)^2 + \eta^2 \bigr]. 
\label{sdbrck}
\end{equation}
This last result is an order of magnitude smaller than Eq.~\ref{sdbr} 
due in part to the shorter $K^+$ lifetime, and in part to the 
approximation of Eq.~\ref{ffexp}. In particular, 
$p$-wave contributions to $F$ could change this result significantly. 

If we use the values of Ref.\cite{buras} for the charm-quark contribution 
with QCD corrections to Eq.~\ref{shortd}, and take $\Lambda_{QCD}=200$~MeV, 
$m_c=1.4$~GeV and $m_t=175$~GeV, we find:
\begin{eqnarray}
B(K_L \to \pi^+ \pi^- \nu \overline{\nu})&\approx& \bigl[1.8 (1.37-\rho)^2 
+0.3 \eta^2 \bigr] \times 10^{-13} \nonumber \\
B(K_L \to \pi^0 \pi^0 \nu \overline{\nu})&\approx& 1 \times 10^{-13}\: 
(1.37-\rho)^2 \nonumber \\
B(K^+ \to \pi^+ \pi^0 \nu \overline{\nu})&\approx& 7 \times 10^{-15} 
\: \bigl[(1.37-\rho)^2 +\eta^2\bigr] \nonumber \\
A_{CP}&\approx&1 \times 10^{-14}\: \eta(1.37-\rho)
\label{numbers}
\end{eqnarray}

Schematically, the decay $K_L \to \pi^+ \pi^- \nu \overline{\nu}$ is induced 
by the operator of Eq.~\ref{shortd} through diagrams such as those in 
Figs.~\ref{diagrams}a and~\ref{diagrams}b. In these two diagrams the 
short-distance four-fermion operator of Eq.~\ref{shortd} is represented 
by the full crossed circle. Fig.~\ref{diagrams}a represents constant 
form factors and appears at lowest order in $\chi PT$, whereas contributions 
such as the one depicted in Fig.~\ref{diagrams}b introduce momentum 
dependence into the form factors and arise at higher orders in $\chi PT$. 

There are also long-distance contributions to the decays $K_L \to \pi^+ 
\pi^- \nu \overline{\nu}$ and we have shown some of them in 
Fig.~\ref{diagrams}c-f. 
Fig.~\ref{diagrams}c represents a charged weak current followed by a neutral 
weak current interaction. There are several such contributions: an eta 
pole can replace the pion pole; there can be higher order momentum dependence 
introduced as in Fig.~\ref{diagrams}d; the neutral current interaction can 
occur in the kaon leg as in Fig.~\ref{diagrams}e and so on. It is easy to 
see that these contributions are much smaller than the short distance 
contribution Eq.~\ref{sdbr}. For example, the diagram in Fig.~\ref{diagrams}c 
gives at lowest order in $\chi PT$ a contribution equivalent to having a 
form factor $H \approx 0.03$ in Eq.~\ref{klamp}, 
much smaller than the corresponding short distance factor 
$h(0) (\rho^0_\ell -\rho)$. 
The lepton pole diagrams in Fig.~\ref{diagrams}f are also found to give 
a very small correction to the rate. After summing over the three leptons 
their contribution is: 
$B_{\ell}(K_L \to \pi^+ \pi^- \nu \overline{\nu}) = 6.9 \times 10^{-17}$.

\setlength{\unitlength}{0.0125in}%
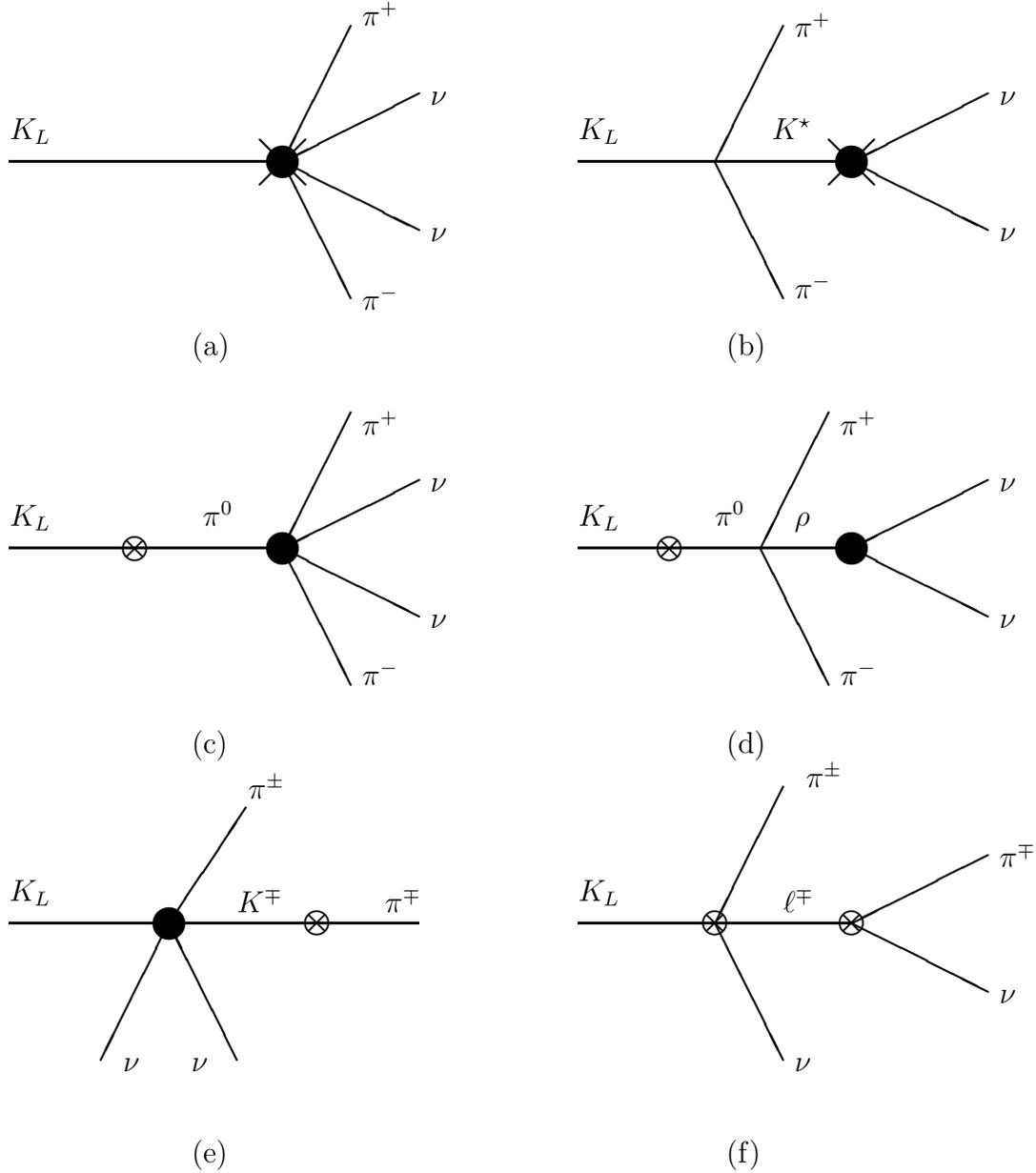
\begin{figure}[htb]
\begin{picture}(435,511)(145,270)
\thicklines
\put(145,710){\line( 1, 0){120}}
\put(485,770){\line(-1,-2){ 30}}
\put(455,710){\line( 1,-2){ 30}}
\put(575,740){\line(-2,-1){ 60}}
\put(515,710){\line( 2,-1){ 60}}
\put(395,710){\line( 1, 0){120}}
\put(505,720){\line( 1,-1){ 20}}
\put(525,720){\line(-1,-1){ 20}}
\put(515,710){\circle*{14}}
\put(215,375){\circle*{14}}
\put(245,315){\line(-1, 2){ 30}}
\put(215,375){\line(-1,-2){ 30}}
\put(325,570){\line(-2,-1){ 60}}
\put(265,540){\line( 2,-1){ 60}}
\put(145,540){\line( 1, 0){120}}
\put(295,600){\line(-1,-2){ 30}}
\put(265,540){\line( 1,-2){ 30}}
\put(265,540){\circle*{14}}
\put(200,540){\makebox(0,0){$\bigotimes$}}
\put(485,435){\line(-1,-2){ 30}}
\put(455,375){\line( 1,-2){ 30}}
\put(255,720){\line( 1,-1){ 20}}
\put(275,720){\line(-1,-1){ 20}}
\put(265,710){\circle*{14}}
\put(575,405){\line(-2,-1){ 60}}
\put(515,375){\line( 2,-1){ 60}}
\put(575,570){\line(-2,-1){ 60}}
\put(515,540){\line( 2,-1){ 60}}
\put(395,540){\line( 1, 0){120}}
\put(505,600){\line(-1,-2){ 30}}
\put(475,540){\line( 1,-2){ 30}}
\put(515,540){\circle*{14}}
\put(215,375){\line( 2, 3){ 33.846}}
\put(145,375){\line( 1, 0){180}}
\put(280,375){\makebox(0,0){$\bigotimes$}}
\put(455,375){\makebox(0,0){$\bigotimes$}}
\put(515,375){\makebox(0,0){$\bigotimes$}}
\put(295,770){\line(-1,-2){ 30}}
\put(265,710){\line( 1,-2){ 30}}
\put(325,740){\line(-2,-1){ 60}}
\put(265,710){\line( 2,-1){ 60}}
\put(395,375){\line( 1, 0){120}}
\put(435,540){\makebox(0,0){$\bigotimes$}}
\put(330,735){\makebox(0,0)[lb]{\raisebox{0pt}[0pt][0pt]{$\nu$}}}
\put(330,675){\makebox(0,0)[lb]{\raisebox{0pt}[0pt][0pt]{$\nu$}}}
\put(580,735){\makebox(0,0)[lb]{\raisebox{0pt}[0pt][0pt]{$\nu$}}}
\put(580,675){\makebox(0,0)[lb]{\raisebox{0pt}[0pt][0pt]{$\nu$}}}
\put(580,340){\makebox(0,0)[lb]{\raisebox{0pt}[0pt][0pt]{$\nu$}}}
\put(490,310){\makebox(0,0)[lb]{\raisebox{0pt}[0pt][0pt]{$\nu$}}}
\put(580,400){\makebox(0,0)[lb]{\raisebox{0pt}[0pt][0pt]{$\pi^\mp$}}}
\put(310,380){\makebox(0,0)[lb]{\raisebox{0pt}[0pt][0pt]{$\pi^\mp$}}}
\put(195,310){\makebox(0,0)[lb]{\raisebox{0pt}[0pt][0pt]{$\nu$}}}
\put(225,310){\makebox(0,0)[lb]{\raisebox{0pt}[0pt][0pt]{$\nu$}}}
\put(245,380){\makebox(0,0)[lb]{\raisebox{0pt}[0pt][0pt]{$K^\mp$}}}
\put(230,550){\makebox(0,0)[lb]{\raisebox{0pt}[0pt][0pt]{$\pi^0$}}}
\put(330,565){\makebox(0,0)[lb]{\raisebox{0pt}[0pt][0pt]{$\nu$}}}
\put(330,505){\makebox(0,0)[lb]{\raisebox{0pt}[0pt][0pt]{$\nu$}}}
\put(455,550){\makebox(0,0)[lb]{\raisebox{0pt}[0pt][0pt]{$\pi^0$}}}
\put(580,565){\makebox(0,0)[lb]{\raisebox{0pt}[0pt][0pt]{$\nu$}}}
\put(580,505){\makebox(0,0)[lb]{\raisebox{0pt}[0pt][0pt]{$\nu$}}}
\put(490,550){\makebox(0,0)[lb]{\raisebox{0pt}[0pt][0pt]{$\rho$}}}
\put(145,385){\makebox(0,0)[lb]{\raisebox{0pt}[0pt][0pt]{$K_L$}}}
\put(395,720){\makebox(0,0)[lb]{\raisebox{0pt}[0pt][0pt]{$K_L$}}}
\put(395,550){\makebox(0,0)[lb]{\raisebox{0pt}[0pt][0pt]{$K_L$}}}
\put(395,385){\makebox(0,0)[lb]{\raisebox{0pt}[0pt][0pt]{$K_L$}}}
\put(145,550){\makebox(0,0)[lb]{\raisebox{0pt}[0pt][0pt]{$K_L$}}}
\put(145,720){\makebox(0,0)[lb]{\raisebox{0pt}[0pt][0pt]{$K_L$}}}
\put(300,770){\makebox(0,0)[lb]{\raisebox{0pt}[0pt][0pt]{$\pi^+$}}}
\put(300,645){\makebox(0,0)[lb]{\raisebox{0pt}[0pt][0pt]{$\pi^-$}}}
\put(300,590){\makebox(0,0)[lb]{\raisebox{0pt}[0pt][0pt]{$\pi^+$}}}
\put(300,480){\makebox(0,0)[lb]{\raisebox{0pt}[0pt][0pt]{$\pi^-$}}}
\put(490,765){\makebox(0,0)[lb]{\raisebox{0pt}[0pt][0pt]{$\pi^+$}}}
\put(490,650){\makebox(0,0)[lb]{\raisebox{0pt}[0pt][0pt]{$\pi^-$}}}
\put(510,590){\makebox(0,0)[lb]{\raisebox{0pt}[0pt][0pt]{$\pi^+$}}}
\put(510,480){\makebox(0,0)[lb]{\raisebox{0pt}[0pt][0pt]{$\pi^-$}}}
\put(485,380){\makebox(0,0)[lb]{\raisebox{0pt}[0pt][0pt]{$\ell^\mp$}}}
\put(480,720){\makebox(0,0)[lb]{\raisebox{0pt}[0pt][0pt]{$K^\star$}}}
\put(250,430){\makebox(0,0)[lb]{\raisebox{0pt}[0pt][0pt]{$\pi^\pm$}}}
\put(495,435){\makebox(0,0)[lb]{\raisebox{0pt}[0pt][0pt]{$\pi^\pm$}}}
\put(225,625){\makebox(0,0)[lb]{\raisebox{0pt}[0pt][0pt]{(a)}}}
\put(225,450){\makebox(0,0)[lb]{\raisebox{0pt}[0pt][0pt]{(c)}}}
\put(225,270){\makebox(0,0)[lb]{\raisebox{0pt}[0pt][0pt]{(e)}}}
\put(460,625){\makebox(0,0)[lb]{\raisebox{0pt}[0pt][0pt]{(b)}}}
\put(460,450){\makebox(0,0)[lb]{\raisebox{0pt}[0pt][0pt]{(d)}}}
\put(460,270){\makebox(0,0)[lb]{\raisebox{0pt}[0pt][0pt]{(f)}}}
\end{picture}
\caption{Classes of diagrams that contribute to the decay $K_L \to \pi^+
\pi^- \nu \overline\nu$: (a) Short distance vertex. 
(b) Higher order $\chi PT$ corrections to (a). (c) Long distance 
contribution from a charged current weak interaction followed by a neutral 
current weak interaction. (d) Higher 
order corrections to (c). (e) Long distance contribution from a neutral 
current weak interaction followed by a charged current 
weak interaction. (f) Long distance lepton-pole 
contribution from two charged current weak interactions.}
\label{diagrams}
\end{figure}

\begin{figure}[htb]
\centerline{\epsfxsize=4.5in\epsfbox{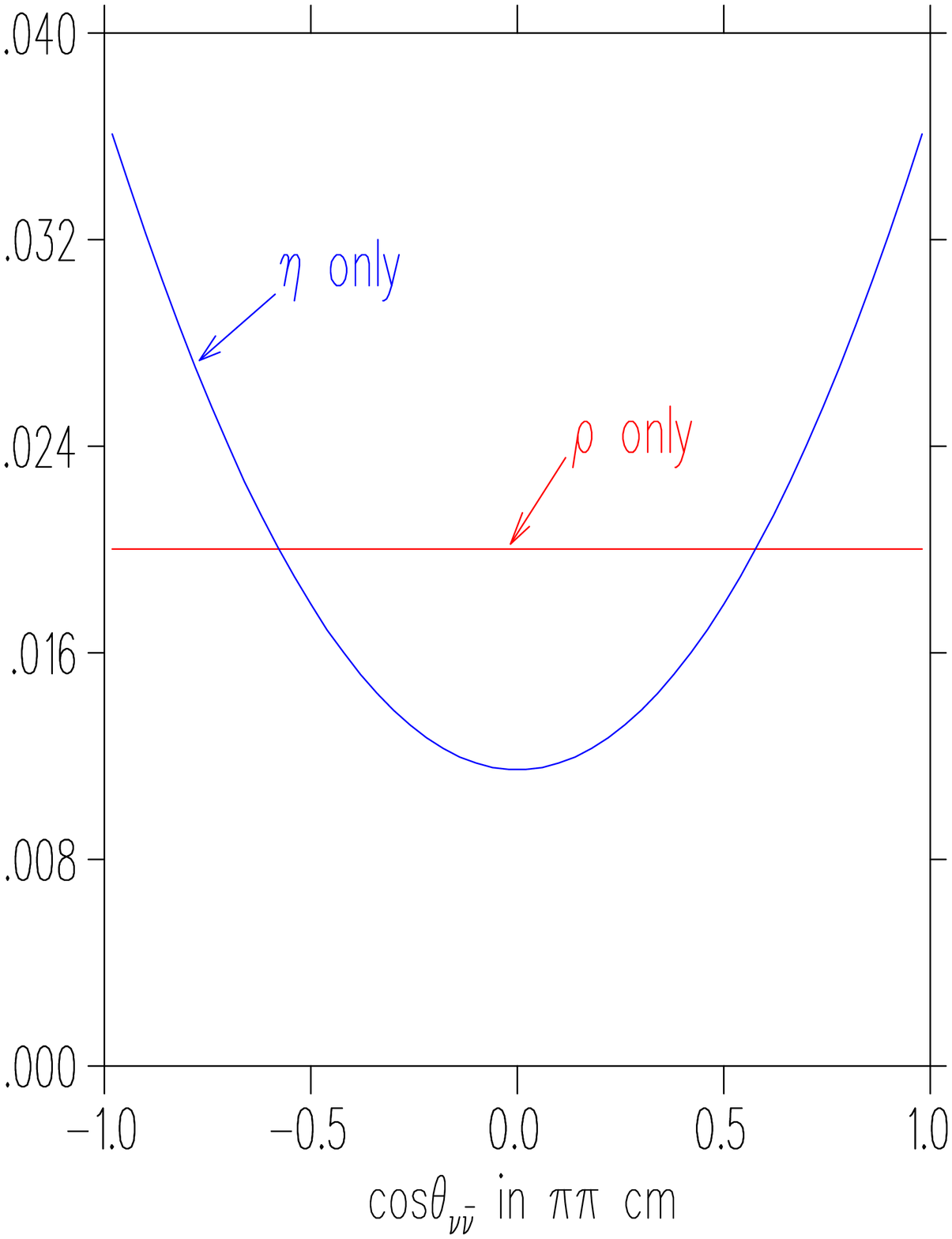}}
\caption{$d\Gamma (K_L \to \pi^+ \pi^- \nu \overline{\nu})/ 
d\cos \theta_{\nu\bar\nu}^{\pi \pi}$. $\theta_{\nu\bar\nu}^{\pi \pi}$
is the angle between the $\pi^+$ momentum and the sum of $\nu$ and $\bar \nu$
momenta evaluated in the $\pi^+ \pi^-$ center of mass.
The two terms corresponding to 
the $F^2$ (marked as $\rho$-only)  and $G^2$ (marked as $\eta$-only) 
contributions are shown. The distribution given by each term has been 
normalized to unit area.}
\label{cos}
\end{figure}

It is amusing to note that because of the different angular momentum
characteristics of the terms involving $\rho$ and $\eta$, in
principle these quantities could be separately extracted from a
sufficiently large sample of  $K_L \to \pi^+ \pi^- \nu \overline\nu$.
An indication of this can be seen in 
Fig.~\ref{cos}, which shows the contrasting dependences on 
$\cos \theta_{\nu\bar\nu}^{\pi \pi}$ of the term proportional to
$(\rho^0_\ell -\rho)^2$ and that proportional to $\eta^2$.  Here
$\theta_{\nu\bar\nu}^{\pi \pi}$ is the angle between the
$\pi^+$ and vector sum of the $\nu$ and $\bar\nu$ momenta in
the $\pi-\pi$ cm system.  In practice, however, the relatively small size of
the $\eta$ contribution will make it very hard to extract.  Thus
this process will mainly serve to determine a value for $\rho$.

It is also worth pointing out that the rare decay modes we discuss in 
this note, $K \rightarrow \pi\pi\nu\overline\nu$, are complementary 
to the decay modes $K \rightarrow \pi\nu\overline\nu$ in searches for 
new physics. This is similar to the complementarity of $K_L \rightarrow 
\mu^\pm e^\mp$ and $K \rightarrow \pi \mu^\pm e^\mp$ in searches for 
lepton flavor violating interactions. The modes with one pion in the final 
state are only sensitive to new interactions inducing vector or scalar 
quark currents, whereas the modes with two pions in the final state are also 
sensitive to axial-vector and pseudo-scalar quark currents.

The detection of $K_L \to \pi^+ \pi^- \nu \overline\nu$ will represent
a major experimental challenge, particularly from the point of view
of background rejection.  The expected size of the branching ratio
is only about an order of magnitude below the current state of the
art (experiments presently running at the BNL AGS are designed to
achieve a sensitivity of $10^{-12}$/event~\cite{bnl}).  It is quite
probable that a supply of $K_L$ sufficient to measure this process
will be available within a few years.   However, distinguishing this 
process from a number of much more copious $K_L$ decays may require 
substantial improvements in present-day photon vetoing and particle 
identification technology.
Table~\ref{t; bg} shows four obvious background possibilities. 

\begin{table}[htb]
\centering
\caption{Some possible backgrounds to $K_L \to \pi^+ \pi^- \nu \overline\nu$}
\begin{tabular}{|c|c|c|c|c|} \hline
& &rejection&rejection&kinematic \\
Mode & BR~\cite{PDB} & technique & factor & rejection needed \\ \hline
$K_L \to \pi^+ \pi^- \pi^0$ & $0.1238$& photon veto & $(10^4)^2$ 
& $1.2 \times 10^4$ \\ 
$K_L \to \pi^+ \pi^- \gamma$ & $4.61 \times 10^{-5}$& photon veto & $10^4$ & 
$4.6 \times 10^4$ \\
$K_L \to \pi^{\mp} \mu^{\pm} \stackrel{(-)} \nu$ & $0.27$& $\mu$ ID & $10^6$ &
 $2.7 \times 10^6$ \\ 
$K_L \to \pi^{\mp} e^{\pm} \stackrel{(-)} \nu$ & $0.387$& e ID & $10^6$ &
 $3.9 \times 10^6$ \\ \hline
\hline
\end{tabular}
\label{t; bg}
\end{table}

\begin{figure}[htb]
\centerline{\epsfxsize=4.5in\epsfbox{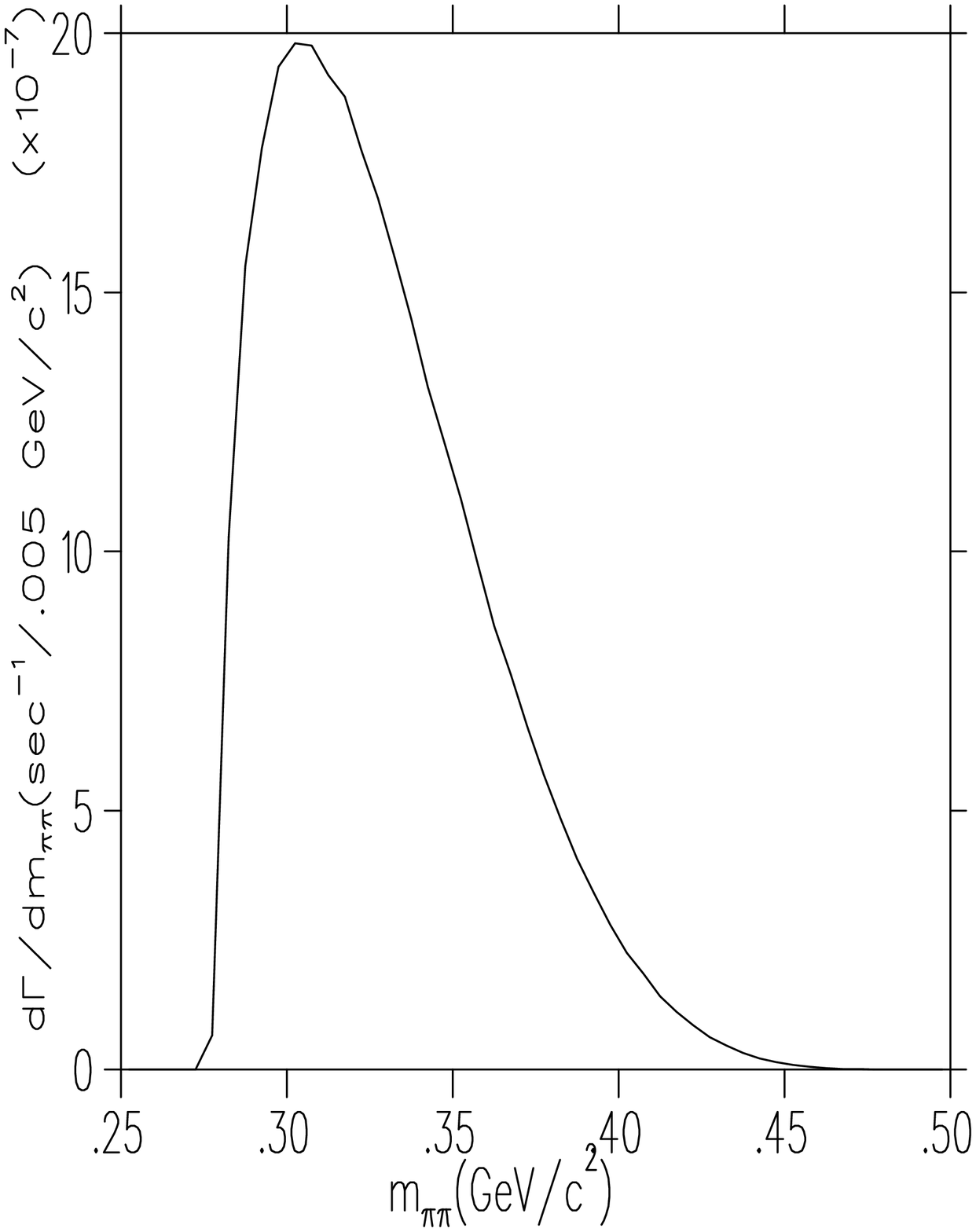}}
\caption{$d\Gamma(K_L\to \pi^+ \pi^- \nu \overline{\nu})/dm_{\pi\pi}$ for
$\rho= -0.05$, $\eta=0.3$, and $A = 0.85$.  The shape of this distribution
is quite insensitive to the values of these parameters over their presently
allowed range.}
\label{invmass}
\end{figure}

The particle ID and photon veto rejections listed are optimistic, 
but not out of the question
for a next-generation experiment.  The optimism lies less in the absolute 
rejections than in the notion that all can be achieved simultaneously 
in the same apparatus.  For each background a very large factor of additional
rejection is required to get to the $\sim 10^{-13}$ level.  These
would have to be supplied via kinematical separation.  In Fig.~\ref{invmass} 
we show the differential distribution $d\Gamma(K_L\to \pi^+ \pi^- \nu 
\overline{\nu})/dm_{\pi\pi}$ for typical values of $\rho$ 
and $\eta$. The shape of this distribution does not change significantly 
when we vary $\rho$ and $\eta$ over their presently allowed range. 
This distribution differs markedly from the corresponding ones in the
background reactions listed in Table~\ref{t; bg}, but not to the extent 
that would allow the rejection factors listed in the rightmost column
to be achieved.  If one adds information on the $K_L$ direction,
variables such as $P02$\cite{po2} can distinguish $K_L\to \pi^+ \pi^- \nu 
\overline{\nu}$ from $K_L \to \pi^+ \pi^- \pi^0$, but are much less
effective against $K_L\to \pi^+ \pi^- \gamma$ and $K \ell 3$.
To obtain really large rejections, it will be necessary to 
to determine the $K_L$ momentum.  Then, for each of the backgrounds in
Table~\ref{t; bg}, one  can compute a missing mass recoiling from the
charged system that should be a value unique to that background\cite{kl3}.
Unfortunately, the $K_L$ momentum can only be accurately 
measured when it is rather low ($\leq 2$ GeV/c), whereas photon
vetoing tends to be more effective at higher
momenta. 

Detection of $K_L \to \pi^0 \pi^0 \nu\bar\nu$ is likely to be
even more challenging, because of the relative difficulty in reconstructing
all-neutral final states, and because $B(K_L \to 3 \pi^0)/B(K_L \to \pi^0 
\pi^0 \nu\bar\nu) \approx 3 \times B(K_L \to \pi^+ \pi^- \pi^0)/B(K_L \to 
\pi^+ \pi^- \nu\bar\nu)$.  However there are also some advantages in the 
neutral case.  One does not need to compromise acceptance and photon vetoing 
power by accommodating magnetic reconstruction and charged particle 
identification.  What is more, certain backgrounds, such as $K_L \to \pi^0
\pi^0 \gamma$, are much smaller than their charged analogues. 
It would be natural to add this mode to the menu of any experiment aimed
at detecting $K_L \to \pi^0 \nu\bar\nu$, if the trigger rate allows.

In conclusion we have proposed a new, theoretically clean, way of probing 
the CKM parameter $\rho$. 
This should serve as an additional motivation for a new analysis 
of $K_{\ell4}$ decays with a more detailed study of the form-factors.

After completion of this work we became aware of Ref.\cite{geng} which studies
the reaction $K_L \rightarrow \pi^+ \pi^- \nu \overline{\nu}$ using chiral
perturbation theory and obtains results similar to ours. Our calculation
differs from that in Ref.\cite{geng} in that we obtain the matrix elements
directly from the form factors measured in $K_{\ell 4}$ using isospin
symmetry. Our results are also presented in a way that we find more
illuminating than that used by Ref.\cite{geng}. Ref.\cite{geng} obtains
allowed ranges for the CKM angles from fits to other processes and
presents final results for the rate of
$K_L \rightarrow \pi^+ \pi^- \nu \overline{\nu}$ based on those fits.
Instead, we present
simple numerical results in terms of the CKM angles that can be easily
adapted to changing constraints on the values of the CKM parameters.
We also discuss two additional modes,
$K_L \rightarrow \pi^0 \pi^0 \nu \overline{\nu}$ and
$K^+ \rightarrow \pi^+ \pi^0 \nu \overline{\nu}$,
as well as a possible $CP$-odd observable that are not studied
in Ref.\cite{geng}.


The work of L.L. was supported by DOE contract No. DE-AC02-76CH00016 and the 
work of G.V. was supported in part by the DOE OJI program under contract number
DE-FG02-92ER40730. We thank John Donoghue, Mark Ito and William Marciano 
for useful discussions.



\begin{thebibliography}{999}
\bibitem{cptinv}{We assume CPT invariance and we drop a very 
small contribution from $CP$ violation in the $K^0-\overline{K}^0$ mass 
matrix.}
\bibitem{po2}{This commonly used variable is defined as
$P02 \equiv [(m_K^2 - m_{\pi^0}^2 - m_{\pi\pi}^2)^2 - 4  
m_{\pi^0}^2 m_{\pi\pi}^2 - 4 m_K^2 (P_{\pi\pi}^T )^2 ]
/[ (P_{\pi\pi}^T )^2 + m_{\pi\pi}^2 ]$, where $m_{\pi\pi}$ is
the effective mass of the charged $\pi$ system and $P_{\pi\pi}^T$ is
the component of the vector sum of the pion momenta transverse to
the $K_L$ direction.}
\bibitem{kl3}{To identify $K\ell3$, one must assign the lepton mass to each 
charged track in turn; one combination will give missing mass $\approx 0$.}
\end{thebibliography}
\end{document}